\documentclass[superscriptaddress,twocolumn,aps,floatfix]{revtex4-2}

\usepackage[colorlinks=true,citecolor=blue,linkcolor=magenta]{hyperref}
\usepackage{changes}
\usepackage{lineno}
\usepackage{stmaryrd}
\usepackage{bm}
\usepackage{graphicx}
\usepackage{epsfig}
\usepackage{color}
\usepackage{amssymb}
\usepackage{amsmath}
\usepackage{array}
\usepackage{braket}
\usepackage{soul}
\usepackage{cleveref}
\usepackage[caption=false]{subfig}
\usepackage[english]{babel}
\usepackage{letltxmacro}
\usepackage{siunitx}
\usepackage{dsfont}
\usepackage{ulem}

\usepackage{enumitem}
\setitemize{noitemsep,topsep=0pt,parsep=0pt,partopsep=0pt}

\newcommand{\MPINAT}{Department of Ultrafast Dynamics, Max Planck Institute for Multidisciplinary Sciences, D-37077 G\"{o}ttingen, Germany}
\newcommand{\GOE}{4th Physical Institute, Georg-August-Universit\"{a}t G\"{o}ttingen, D-37077 G\"{o}ttingen, Germany}
\newcommand{\EPFL}{Swiss Federal Institute of Technology Lausanne (EPFL), CH-1015 Lausanne, Switzerland}
\newcommand{\CfQS}{Center for Quantum Science and Engineering, EPFL, CH-1015 Lausanne, Switzerland}

\usepackage{scalerel}
\usepackage{tikz}
\usetikzlibrary{svg.path}

\usepackage{hyperref} %<--- Load after everything else

\begin{document}

\title{Electrons herald non-classical light}
\author{Germaine Arend}
\thanks{These authors contributed equally.}
\affiliation{\MPINAT}
\affiliation{\GOE}
\author{Guanhao Huang}
\thanks{These authors contributed equally.}
\affiliation{\EPFL}
\affiliation{\CfQS}
\author{Armin Feist}
\email{armin.feist@mpinat.mpg.de}
\affiliation{\MPINAT}
\affiliation{\GOE}
\author{Yujia Yang}
\email{yujia.yang@epfl.ch}
\affiliation{\EPFL}
\affiliation{\CfQS}
\author{Jan-Wilke Henke}
\affiliation{\MPINAT}
\affiliation{\GOE}
\author{Zheru Qiu}
\affiliation{\EPFL}
\affiliation{\CfQS}
\author{Hao Jeng}
\affiliation{\MPINAT}
\affiliation{\GOE}
\author{Arslan Sajid Raja}
\affiliation{\EPFL}
\affiliation{\CfQS}
\author{Rudolf Haindl}
\affiliation{\MPINAT}
\affiliation{\GOE}
\author{Rui Ning Wang}
\affiliation{\EPFL}
\affiliation{\CfQS}
\author{Tobias J. Kippenberg}
\email{tobias.kippenberg@epfl.ch}
\affiliation{\EPFL}
\affiliation{\CfQS}
\author{Claus Ropers}
\email{claus.ropers@mpinat.mpg.de}
\affiliation{\MPINAT}
\affiliation{\GOE}

\maketitle
\date{\today}

\section*{Abstract}
\textbf{Free electrons are a widespread and universal source of electromagnetic fields. The past decades witnessed ever-growing control over many aspects of electron-generated radiation, from the incoherent emission produced by X-ray tubes to the exceptional brilliance of free-electron lasers. Reduced to the elementary process of quantized energy exchange between individual electrons and the electromagnetic field, electron beams may facilitate future sources of tunable quantum light. However, the quantum features of such radiation are tied to the correlation of the particles, calling for the joint electronic and photonic state to be explored for further applications. Here, we demonstrate the coherent parametric generation of non-classical states of light by free electrons. We show that the quantized electron energy loss heralds the number of photons generated in a dielectric waveguide. In Hanbury-Brown-Twiss measurements, an electron-heralded single-photon state is revealed via antibunching intensity correlations, while two-quantum energy losses of individual electrons yield pronounced two-photon coincidences. The approach facilitates the tailored preparation of higher-number Fock and other optical quantum states based on controlled interactions with free-electron beams.}\\

\textbf{One sentence summary:} We demonstrate free-electron-generated non-classical multiphoton states in a photonic waveguide, leveraging two- and three-particle correlations produced by quantized parametric electron-photon scattering.\\

Quantum states of light facilitate manifold applications in communication~\cite{Ursin2007},
computation~\cite{Kok2007,Zhong2020}, and sensing~\cite{Degen2017,Ganapathy2023}. As a hallmark of quantum optics, photon number states have no classical counterpart and are essential in quantum metrology~\cite{Holland1993} and information processing~\cite{Broome2013, Wang2020}. Furthermore, advanced protocols promise the generation of more exotic states, such as Schrödinger's cat~\cite{Ourjoumtsev2007} and GKP states ~\cite{Gottesman2001} with applications in quantum computing~\cite{Baranes2023, Campagne-Ibarcq2020, Reglade2024}. 
Multiple approaches and platforms have been used for number-state generation, including circuit QED systems~\cite{Geremia2006, Zhou2012, Uria2020} and superconducting structures~\cite{Hofheinz2008} in the microwave regime, as well as spontaneous parametric down-conversion~\cite{Waks2006, Ourjoumtsev2006, Cooper2013, Tiedau2019, Harder2016} in the visible to near-infrared. \\

In the search for new sources of non-classical light, free electrons present an attractive pathway. In pioneering works, electron-excited single quantum emitters, such as color centers and point defects, have already shown antibunching photon statistics in incoherent emission~\cite{Bourrellier2016, Fiedler2023, Tizei2013}. Parametric photon generation by coherent cathodoluminescence~\cite{Vesseur2007, Christopher2020, Brenny2014, Coenen2014, Coenen2017, Sannomiya2020, Scheucher2022a}, on the other hand, promises sophisticated photonic states without involving Fermionic excitations, preserves the quantum phase between electrons and photons~\cite{Kfir2021}, and allows for enhancements by tunable phase matching~\cite{Kfir2020, Muller2021, Feist2022}. In this context, photonic integrated circuits (PICs)~\cite{Wang2020a, Elshaari2020, Liu2021} offer low-loss photon transport, as well as versatile mode tailoring, facilitating efficient scattering between electrons and optical modes~\cite{Henke2021}. Combining free electrons and PICs thus provides a unique platform for the excitation, control, and observation of classical and quantum optical states at widely variable wavelengths.\\

A key element of realizing electron-driven quantum light will be to harness the correlations and entanglement of photonic and electronic states. In particular, electron-photon coincidence detection~\cite{Kruit1984, Graham1986, Bendana2011, Jannis2019, Feist2022, Varkentina2022, Yanagimoto2023} has yielded detailed information on the joint electron-photon state in time, energy, and momentum.
Post-selection of time-correlated electrons and photons was recently shown to improve x-ray elemental mapping~\cite{Jannis2019}, as well as photonic-mode imaging and spectroscopy~\cite{Feist2022, Varkentina2022}. Furthermore, a temporal fingerprint of the interaction may yield material- and excitation-specific decay times~\cite{Varkentina2023, Taleb2024} and improve the signal-to-noise ratio (SNR) in electron microscopy~\cite{Koppell2025, Feist2022}. Coherent cathodoluminescence combined with post-selection is further expected to enable the preparation of electron-heralded Fock~\cite{Kfir2019, BenHayun2021}, GKP~\cite{Dahan2023, Baranes2023}, and other complex quantum states of light~\cite{Dahan2023, Sirotin2024}. 
However, the preparation of non-classical optical states via parametric electron-photon interactions has not yet been shown.\\

Here, we demonstrate the generation of electron-heralded non-classical light using spontaneous parametric scattering in an integrated photonics waveguide. We merge energy-resolved single-electron detection with two-photon coincidence measurements in an optical Hanbury-Brown Twiss (HBT) setup to probe strong correlations between the generated photon number and the quantized electron energy loss. Electron heralding to a single-photon energy change yields prominent photon anti-bunching. The high-contrast detection of threefold coincidences evidences two-photon generation, establishing a novel platform for heralded photon Fock states in integrated photonics-based quantum technology.\\

\begin{figure*}[th]
    \centering
    \includegraphics{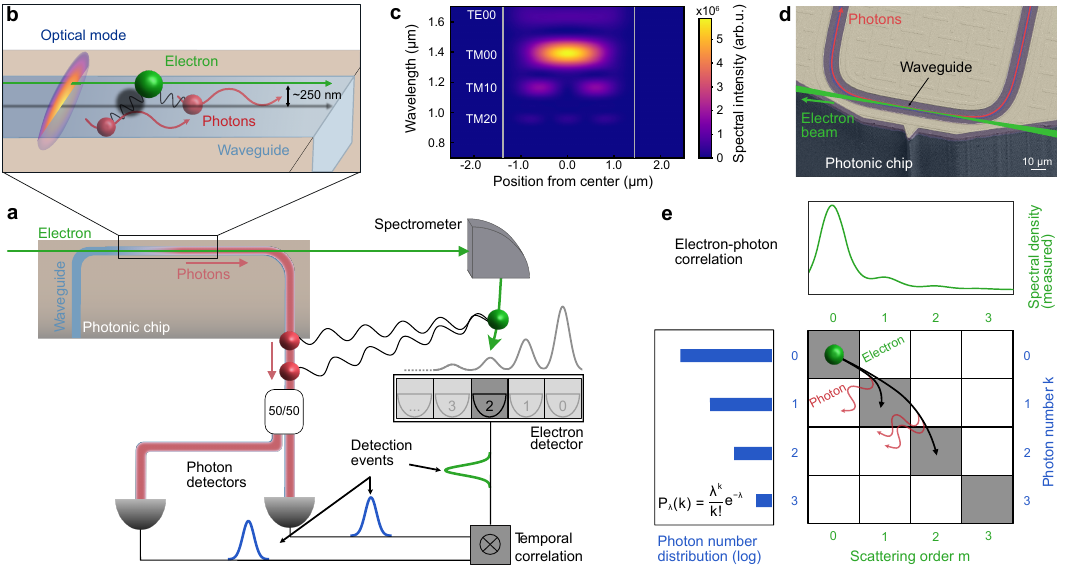}
    \caption{\textbf{Correlation measurement of waveguide-coupled photons and free electrons.} \textbf{a} A beam of electrons passes a dielectric waveguide, generating photons that are guided and detected in a fiber-based HBT setup. The electron and photon arrival times are temporally correlated. \textbf{b} Enlarged sketch of the interaction region: The electron passes the waveguide surface at an impact parameter of around 250~nm to interact with the evanescent vacuum field of the waveguide modes. \textbf{c} Simulation of the photon spectral distribution generated in a $40\,\mu$m straight waveguide by 100-keV electrons (impact parameter 250~nm). Gray vertical lines: waveguide boundaries. \textbf{d} False-color image of an example photonic chip (scanning electron micrograph). The chip surface is covered with ITO (brighter surface: ITO on metal, darker surface: ITO on Si\textsubscript{3}N\textsubscript{4} or on SiO\textsubscript{2}, side walls: SiO\textsubscript{2}) \textbf{e} Schematic of the correlated quantized electron energy loss and photon number. The electron (green) scatters at the optical mode, generating a random number of photons (red). After detection, the signal correlation links the photon number $k$ (vertical, blue) to the electron scattering order $m$ (horizontal, green). The generation process is expected to follow a Poissonian distribution.}
    \label{fig:figure1_setup}
\end{figure*}

In our study, electron-induced photon generation takes place at a $\text{Si}_3\text{N}_4$ waveguide (\SI{2.2}{\micro\metre} $\times$ \SI{780}{\nano\metre} cross-section, geometry as seen in Fig.~\ref{fig:figure1_setup}\textbf{a,d}, embedded in SiO\textsubscript{2})  on a photonic chip, which is inserted into the sample plane of a transmission electron microscope (TEM) (cf. Fig.~\ref{fig:figure1_setup}\textbf{a}). A weakly focused continuous electron beam (100~keV energy, 30~nm beam diameter, 1.1~mrad half angle) passes the waveguide parallel to its surface at an impact parameter (distance to the surface) of about 250~nm (cf. Fig.~\ref{fig:figure1_setup}\textbf{b}). Mediated by the structure's dielectric response, the electrons at initial energy $E_0$ interact with the optical vacuum of the waveguide modes. The interaction generates photons in the modes in a parametric process~\cite{GarciadeAbajo2010, Bendana2011, Feist2022}, characterized by a direct energy exchange between electrons and photons that leaves the dielectric material unexcited. The interaction can, therefore, be described as an inelastic electron-photon scattering which correlates and entangles the final electron energy $E_0-\sum_{i=1}^k\hbar\omega_i$ with $k$ generated photons at frequencies $\omega_i$. In this situation, the frequencies generally lie within a spectral continuum of possible excitations (scattering operator given in the SI)~\cite{Huang2023}, defined by phase-matching between the electron (group) velocity and photon phase velocity~\cite{Kfir2019, Bendana2011}. It is worth noting that this phase-matching is simultaneously satisfied for all photon orders, as the scattering with one or more photons does not significantly alter the velocity of the electron or its path during the passage of the structure (non-recoil condition \cite{GarciadeAbajo2010, Talebi2018b}). \\
In an effective single-mode description, the resulting electron-photon state can be simplified to
\begin{align}
    \ket{\psi_e,\psi_{ph}}= \sum_k c_{k} \ket{E_0-k\hbar\omega_0,k},\label{eq:states}
    \end{align}
Here, the coefficients
    \begin{align}
c_{k}&\approx e^{-|g_0|^2/2}g_0^k/\sqrt{k!},\label{eq:ck}
\end{align}
correspond to a Poissonian process of quantized energy exchanges producing $k$ photons near a central optical frequency $\omega_0$, with a dimensionless coupling constant $g_0 = \sqrt{\int d\omega |g_\omega|^2}$ obtained by integration over the spectral coupling strength density $g_{\omega}$.\\
For the designed structure, the interaction is dominated by the quasi-TM\textsubscript{00} (transverse-magnetic) mode at wavelengths between 1300 and 1400~nm, while other excited optical modes are suppressed (see simulation results in Fig.~\ref{fig:figure1_setup}\textbf{c}).\\

The induced optical state is analyzed by coincidence detection of the generating electrons and the created photons (cf. Fig.~\ref{fig:figure1_setup}\textbf{b}). While the photons are guided through fibers to optical single-particle detectors, the electrons are dispersed in a magnetic spectrometer and recorded with an event-based detector (Timepix3 ASIC), which delivers the kinetic energy and arrival time of each electron. The readout electronics of the electron camera further provide time stamps for the arrival time on both photon detectors for electron-photon-photon temporal correlations.
We expect to detect different photon numbers $k$ in each parametric electron scattering event, with the $k$ photons being temporally linked to electron energy shifts of $m\hbar\omega_0$ (cf. Fig.~\ref{fig:figure1_setup}\textbf{e}).
For an ideal electron-photon state (Eq.~\ref{eq:states}), the photon number and quanta of electron energy change are identical ($k=m$), with probabilities following $c_k^2$. In this case, combining temporal particle correlation and electron energy analysis enables heralding schemes, in which the detection of an electron in a specific energy state $\ket{E_0-k\hbar\omega}$ unambiguously heralds the photon number state $\ket{k}$. In practice, finite optical transmission and inelastic scattering that does not inject light into the waveguide will render the quantized electron energy loss $m$ an upper bound $k\leq m,$ which to reach will be a technological target. \\

\begin{figure}[th]
    \centering
   \includegraphics{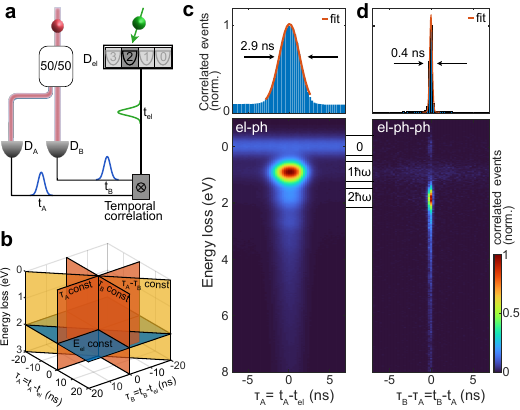}
    \caption{\textbf{Three-particle correlation for one electron and two photons.} \textbf{a}  Schematic of the detection setup with the particle arrival times $t_\textrm{el}, t_\textrm{A}, t_\textrm{B}$ on the fiber-coupled photon detectors D\textsubscript{A}, D\textsubscript{B} and the electron detector D\textsubscript{el} and electron energy $E_\textrm{el}$ as measured parameters. The arrival times are correlated with each other. \textbf{b} Parameter space for threefold events. The data is analyzed through lineouts and integrals along different axes. \textbf{c} Bottom: Histogram of electron-photon-photon coincidence events $N_\textrm{e,A,B}$ as a function of $\tau_\textrm{A}$ and $E_\textrm{el}$, summed over $\tau_\textrm{B}$. Electron-photon scattering causes a coincidence peak at an electron energy shifted downward by around 0.9~eV. Top: The temporal coincidence uncertainty of around 2.9~ns FWHM represents the timing precision of electron detection. \textbf{d} Bottom: Histogram of particle triples as a function of time delay  $t_\textrm{A}-t_\textrm{B}$ and electron energy $E_\textrm{el}$ filtered to true coincidences between a photon A and an electron.
    The coincidences are dominated by electrons that generated 2 photons. Top: The reduced timing uncertainty of 0.4~ns FWHM follows from the better time resolution in photon detection.}
    \label{fig:figure2_correlation}
    \end{figure}

%\section*{Electron-photon-photon correlation}
We gain access to single- and two-photon generation events and characterize the optical state via an electron-heralded optical HBT detection setup shown in Figures~\ref{fig:figure1_setup}\textbf{a} and \ref{fig:figure2_correlation}\textbf{a}. Two single-photon detectors A and B placed behind a 50/50 beam splitter are synchronously time-stamped by the electron detector, and every electron is assigned to the closest photon on either detector, calculating the time delay between them.
The analysis of the resulting electron-photon-photon correlation counts $N_\textrm{e,A,B}$ is conducted as a function of the electron-energy loss $E_\mathrm{el}$ and the relative time delays $\tau_\textrm{A}=t_\textrm{A}-t_\textrm{el}$ and $\tau_\textrm{B}=t_\textrm{B}-t_\textrm{el}$ between the electron and photon detection (arrival times $t_\textrm{el}$, $t_\textrm{A}$ and $t_\textrm{B}$). An event histogram over these degrees of freedom results in a data cube, which we investigate along cuts or in projection on different axes (cf. Fig.~\ref{fig:figure2_correlation}\textbf{b}).   \\

The electron-photon correlation as a function of $\tau_\textrm{A}$ and $E_\textrm{el}$ is given in Fig.~\ref{fig:figure2_correlation}\textbf{c}. The correlation exhibits a strong coincidence peak separated from the uncorrelated, mostly unscattered electrons by approximately 0.9~eV (corresponding to a free-space photon wavelength of $\lambda=1350-1420$~nm). Coincidences are found within a temporal interval of 2.9~ns (FWHM, full-width at half-maximum), mainly defined by the electron detection jitter $\delta t_\textrm{el}$ ~\cite{Feist2022}. Above a constant background in $\tau_\textrm{A}$, the temporally confined coincidence peak consists of events in which at least one electron-induced photon was generated, and both a photon and its generating electron were detected. The detected electron-photon coincidences primarily originate from the excitation of the detectable TM\textsubscript{00} mode. Higher-order modes are suppressed in the chip-fiber coupling, and the weakly-excited TE\textsubscript{00} mode is further diminished due to the detector cutoff wavelength. 

Electron-heralded two-photon generation can be observed via a photon A - photon B correlation trace, filtered to coincidences between photon detector A and an electron.
The arising two-dimensional coincidence histogram in Figure \ref{fig:figure2_correlation}\textbf{d} is dominated by two photons correlated to electrons with about 1.8~eV~=~$2\hbar\omega$ energy loss. The photon-photon coincidence peak has a width of $~$400~ps (FWHM, $\delta t_\textrm{ph}<300$~ps photon detector jitter). Residual coincidence events in the zero-loss region and the first loss sideband stem from false coincidences due to electron or photon loss.

\begin{figure}[th]
    \centering
    \includegraphics{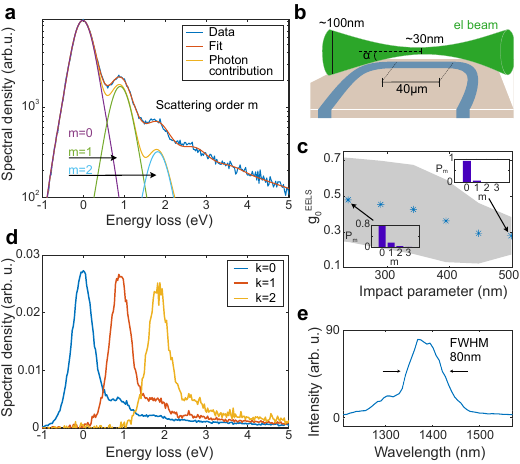}
    \caption{\textbf{Multiple scattering and heralded electron spectra.} \textbf{a} Time-averaged (not photon-correlated) electron spectrum (blue), modeled (red) by cascaded $m$-photon multiple-loss peaks (yellow) and an exponentially decaying continuum. \textbf{b} Sketch of the electron-beam caustic traversing the waveguide with a convergence half-angle of $\alpha=1.1$~mrad, causing some spatial averaging of the coupling constant. \textbf{c} Experimentally observed decay of the mean coupling strength towards larger impact parameters (distance to the waveguide). The standard deviation is given as a gray background. The peak height distributions at different $g^\textrm{EELS}_0$ are included as insets. \textbf{d} Energy distributions of uncorrelated (blue) and correlated electrons (red, yellow) for $k$ detected photons (normalized to the area under the curves). The correlated spectra are obtained by summing over the background-subtracted coincidence peaks shown in Figures \ref{fig:figure2_correlation}\textbf{c,d}. 
    \textbf{e} Measured optical spectrum with a central peak width of about 80~nm, or 65~meV, normalized to the maximum value.
    }
    \label{fig:figure3_spectrum}
\end{figure}

The inelastic electron scattering at the sample is visualized in the electron energy spectra displayed in Figure \ref{fig:figure3_spectrum}. 
The uncorrelated spectrum (c.f. Fig.~\ref{fig:figure3_spectrum}\textbf{a}) contains all channels of inelastic electron scattering~\cite{Egerton2009}, including material-specific radiative and non-radiative effects, quasi-particle excitation~\cite{Schattschneider1987} as well as the desired coherent photon generation into optical modes.
While the electron energy-loss resulting from electron scattering at a dielectric material can be described by a continuous exponential decay in energy~\cite{GarciadeAbajo2010}, phase-matched parametric photon generation leads to the formation of discrete sidebands, downshifted by multiples $m$ of the generated photon's energy $\hbar\omega$ and following a population distribution described by $P_m=|c_m|^2$ (see Equation \ref{eq:ck} with $k=m$ and Fig.~\ref{fig:figure1_setup}\textbf{e}). As the electron scattering channels at the material and into the optical modes are independent of each other, the measured electron spectrum can be modeled by a convolution of the spectra resulting from broadband scattering and from coherent cathodoluminescence in guided optical modes (cf. SI and Fig.~\ref{fig:figure3_spectrum}\textbf{a}, red).\\
Removing the effect of dielectric losses, the spectral component governing coherent photon generation is given by the yellow curve with a peak-to-peak distance between the sidebands close to 0.9~eV. 
The overall coupling constant of the electrons to the waveguide optical modes $g^{\textrm{EELS}}_0$ can be calculated from the population distribution of the individual sidebands $P_m$ (c.f. Fig.~\ref{fig:figure3_spectrum}\textbf{a}, violet to light blue peaks). In the experiment, some averaging of the coupling constant is observed, caused by a broad distribution of impact parameters due to a nonzero convergence angle and finite beam-focus diameter (cf. Fig.~\ref{fig:figure3_spectrum}\textbf{b,c}), as well as time-dependent beam shifts. Using the population ratio between the first three sidebands in the electron spectra, we obtain a coupling constant of $\langle g^\textrm{EELS}_0\rangle = 0.32$ with a deviation of $\Delta_{g^\textrm{EELS}_0} = 0.24$ (see SI for calculations).

Whereas the unheralded electron spectrum includes all inelastic scattering pathways, access to the properties of the detected optical mode is enabled through electron-photon coincidences. The spectra of electrons in coincidence with one (red) or two (yellow) detected photons (cf. Fig.~\ref{fig:figure3_spectrum}\textbf{d}) closely resemble copies of the uncorrelated spectrum (blue) when normalized to the area under the curve. The central wavelength of the TM\textsubscript{00} mode, which dominates the transmitted photons, defines the relative shift of the spectra to be around 0.9~eV. A moderate spectral broadening of 65~meV per electron scattering order can be attributed to the phase-matching bandwidth of 80 nm with the optical mode (cf. Fig.~\ref{fig:figure3_spectrum}\textbf{e}). This corresponds to an effective interaction length of around $40~\mu\text{m}$, in which the electron beam and the optical mode’s field can be regarded as parallel (see simulation results in the SI).\\

While the previously calculated coupling constant $g_0^\textrm{EELS}$ includes all excitable modes, we are primarily interested in the strong coupling to a single optical mode. We gain access to its coupling constant via the intrinsic mode filtering of the optical setup and subsequent electron-photon coincidence detection. 
The coupling strength $g_0$ of electrons to the detected TM\textsubscript{00} mode is determined taking into account optical losses (see SI for transmission measurements). The detection probabilities of one or two photons $P_1$ and $P_2$ are deduced from the area of the $m=1$ and $m=2$ sidebands in the electron-photon and electron-photon-photon coincidence spectra, respectively. They define the number of single- (two-)photon excitation events in which one (two) photons were detected. The calculated $g_0>0.2$ is somewhat lower than the value extracted from the electron spectrum, illustrating the necessity of coincident photon detection for disentangling multiple scattering pathways and, in particular, for isolating coherent photon generation. \\

Next to the spectral characterization of heralded and unheralded electronic states, we study the electron-photon-photon coincidences and their intensity correlations $g^{(2)}$.
By filtering electron-photon coincidence events to electron energies of $E_0 - m\hbar\omega$, the generated optical state is expected to contain $m$ photons, making the heralded state non-classical. The photon statistics of the measured optical state are obtained via (heralded) intensity correlation $g^{(2)}$. \\

\begin{figure}[t]
    \centering
    \includegraphics{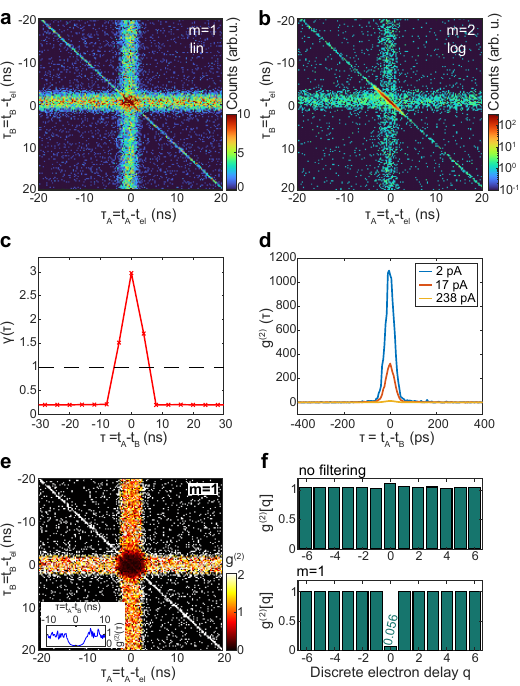}
    \caption{\textbf{Statistical analysis of generated photons.}\\
   \textbf{a,b} Electron-photon-photon correlation events $N_\textrm{e,A,B}$ as a function of photon-electron time delays, selected by electron energy: \textbf{a} scattering order $m=1$, and \textbf{b} scattering order $m=2$. \textbf{c} Violation of the Cauchy-Schwartz Inequality ($\gamma > 1$) for the electron-photon interaction at different time delays $\tau$ (points connected for visibility). \textbf{d} Unheralded photon-photon intensity correlation for varying electron current. \textbf{e} Electron-heralded photon-photon intensity correlation filtered to the energy region $m=1$ as a function of both time delays. Inset: Time-averaged $g^{(2)}$ for $m=1$.
   \textbf{f} Photon-photon intensity correlation as a function of the number of photon-heralding electrons $q$ between the two heralded photons without energy selection (top) and for $m=1$ (bottom).
    }
    \label{fig:figure4_g2}
    \end{figure}

Figures~\ref{fig:figure4_g2}a,b analyze the relative-time histogram $N_\textrm{e,A,B}(\tau_A, \tau_B, E_\textrm{el})$, that counts threefold correlations at delays $\tau_\textrm{A}$ and $\tau_\textrm{B}$ with an electron energy $E_\textrm{el}$. Heralding a photonic state by a specific energy change is achieved by integrating over the corresponding energy windows for single ($m=1$, Fig.~\ref{fig:figure4_g2}\textbf{a}) and two-photon energy loss ($m=2$, Fig.~\ref{fig:figure4_g2}\textbf{b}).
True coincidences of electrons with one photon detector $N_\textrm{e,A$\vee$B}$ (i.e., at detector D\textsubscript{A} OR D\textsubscript{B}) lead to vertical/horizontal bands at $\tau_i\approx0$, while photon-photon coincidences $N_\textrm{A$\wedge$B}$ (excitation of detectors D\textsubscript{A} AND D\textsubscript{B}) with or without a corresponding electron emerge on the diagonal line along $\tau_A-\tau_B\approx0$. Threefold coincidences are found at the intersection of both features (for an in-detail analysis, see SI). While $m=1$ shows electron single-photon coincidences and some photon-photon coincidences not linked to an electron, the plot for $m=2$ is dominated by two photons in coincidence with a detected electron. As the three-particle coincidences are mainly confined to the $2\hbar\omega$ loss region (cf. Figures \ref{fig:figure2_correlation}\textbf{d}, \ref{fig:figure3_spectrum}\textbf{d}), they demonstrate the generation of two-photon states via parametric electron scattering with a two-photon energy loss. The temporal widths of the coincidence features follow the respective detector jitters (cf. Fig.~\ref{fig:figure2_correlation}\textbf{c,d}). Photon-photon coincidences on the diagonal line off the center arise from photon-photon coincidences assigned to a different electron than the generating one, which may have gone undetected.\\

The correlation between the detected electron energy and the generated photon number renders our scheme a versatile heralded photon-number source. The intrinsic heralding efficiency quantifies how the energy-resolved electron detection predicts the number of generated photons $\eta_\textrm{i}^I=N_\textrm{i,j}/(N_\textrm{j}\eta_\textrm{i}^d T_\textrm{i})$~\cite{Signorini2020}. This is the probability that, given a heralding particle $j$ is detected, the heralded particle $i$ also exists. In this measurement setup, the heralding efficiency is calculated for the case of observing $k$ photons when selecting an energy window around a specific electron energy loss $m=1,2$ and vice versa ($\eta_\textrm{i}^dT_\textrm{i}$: transmission losses and detector efficiencies). 
Using the count rates from Figures \ref{fig:figure4_g2}\textbf{a,b}, the efficiencies reach $\eta^I_\textrm{e}>40\%$ for photon-heralded electrons ($m=1,2$) and $\eta^I_{\textrm{A$\vee$B}}=10\%$ and $\eta^I_\textrm{A$\wedge$B}=0.3\%$ for electron-heralded single photons and two photons, respectively. 
Experimentally, $\eta^I_\textrm{A$\wedge$B}$ and $\eta^I_\textrm{A$\vee$B}$ are limited by spectral overlap of the electron-loss peaks (cf. Fig.~\ref{fig:figure3_spectrum}\textbf{a}), which leads to some mixing in the heralding of photon-number states. Further improvements of the heralding efficiency could, thus, include an enhanced energy resolution~\cite{Auad2022}, and a better coupling ideality to the detected mode~\cite{Huang2023}. \\

The quality of a heralded signal is further characterized by the coincidence-to-accidental ratio $\textrm{CAR}= \frac{R_\textrm{sig}-R_\textrm{acc}}{R_\textrm{acc}}$~\cite{Signorini2020}, relating the signal $R_\textrm{sig}$ and background $R_\textrm{acc}$ rates of coincidences in a specific energy loss region $m$. A $\textrm{CAR}$ of $>30$ is measured for single photons in coincidence with an electron at $m=1$, showing good fidelity in generating heralded single photons. The CAR value for electron-heralded two-photon events reaches $>150$ for $m=2$, confirming that detected two-photon states are generated predominantly by double-loss of single electrons and not by random coincidences of two single-photon generation events.\\ 

To test the non-classicality of the radiation process, i.e., whether it is incompatible with classical field theory, we invoke the Cauchy-Schwarz inequality (CSI)~\cite{Reid1986} $\gamma=\frac{g_{E\textrm{A}\vee\textrm{B}}^{(2)}(\tau)}{g_E^{(2)}g^{(2)}(0)}\leq 1$, assuming a classical radiation process. 
The inequality (for derivations, see SI) bounds the temporal cross-correlation between the detected electron energy and photon number $g_{E,\textrm{A$\vee$B}}^{(2)}(\tau)$ to the variation of the electron energy loss $g_E^{(2)}=\frac{\braket{\Delta E_{el}^2}}{\braket{\Delta E_{el}}^2}$ and the temporal intensity correlation of photons in a Hanbury-Brown and Twiss setup $g^{(2)}(\tau)$, normalized to $\tau\gg 0$. For electron-photon coincidence, we find a pronounced violation of the CSI ($\gamma > 1$, cf. Fig.~\ref{fig:figure4_g2}\textbf{c}), which proves the non-classical nature of the process~\cite{Clauser1974, Riedinger2016, Meesala2024}.\\  

The photon statistics of this radiation are further examined by the temporal intensity correlation $g^{(2)}$ of detectors placed in a Hanbury-Brown Twiss (HBT) setup (see Methods). While classical light shows $g^{(2)}(0)\geq g^{(2)}(\tau)$, this is not the case for non-classical light such as single photons or higher photon number states.
The unheralded photon-photon intensity correlation (taken with higher time-resolution, higher-efficiency detectors and a time tagger, see Methods) shows a bunching peak at $\tau=0$~ps~\cite{Yanagimoto2025a}, inversely proportional to the electron current (cf. Fig.~\ref{fig:figure4_g2}\textbf{d}). Following a current-dependent bunching height $g^{(2)}(0)\sim 1+ 1/(I \tau_\mathrm{bin})$, where $I$ is the electron flux rate and $\tau_\mathrm{bin}$ is the temporal data bin width, this behavior is expected when single electrons can create multiple photons, either in a direct process, as demonstrated here, or as a result of cascaded excitation of multiple two-level systems leading to incoherent emission~\cite{Meuret2015, Fiedler2023, Yuge2023, Sola-Garcia2021a}.\\

The statistics of the non-classical electron-heralded photon state are analyzed using the heralded intensity correlation (see Methods)
\begin{align}
g^{(2)}_H(\tau_\textrm{A},\tau_\textrm{B},E_\textrm{el})=\frac{N_\textrm{e,A,B}(\tau_\textrm{A},\tau_\textrm{B},E_\textrm{el})}{N_\textrm{e,A}(\tau_\textrm{A},E_\textrm{el})N_\textrm{e,B}(\tau_\textrm{B},E_\textrm{el})}N_\textrm{e}(E_\textrm{el}), \label{eq:g2-2D}
\end{align}
in which the detected electrons are energy-filtered to a defined loss peak and used as heralds for the photon state. Thus, the correlation equation relates the occurrence of electron-photon-photon coincidences $N_\textrm{e,A,B}$ (cf. Fig.~\ref{fig:figure4_g2}\textbf{a,b}) to their single-photon equivalents $N_\textrm{e,A/B}$ and the number of heralding electrons $N_\textrm{e}$, normalizing $g^{(2)}_H(\tau_\textrm{A},\tau_\textrm{B})$ in this manner to single-photon events ($\tau_{A/B}\approx0$, $\tau_{B/A}\neq 0$). This uncovers the true threefold coincidences at $\tau_\textrm{A$\wedge$B}=0$.\\
Figure \ref{fig:figure4_g2}\textbf{e} displays $g^{(2)}_H(\tau_\textrm{A},\tau_\textrm{B})$ for single-photon electron-energy losses (${m=1}$), as well as the so-called time-averaged heralded intensity correlation $\overline{g}_H^{(2)}(\tau)$ reduced to the relative photon delay (inset; see SI for calculation)~\cite{Bocquillon2009, Bettelli2010}. The background level and the diagonal feature ($\tau_\textrm{A$\wedge$B} \neq 0$) originate from accidental coincidences (cf. Fig.~\ref{fig:figure4_g2}\textbf{a}). At $\tau_\textrm{A$\wedge$B}\approx 0$, $g^{(2)}_H$ drops sharply to near-zero values, as also visible in the inset. Hence, for $m=1$, two photons detected at similar times are generated by two different electrons. Usually, accidental and true heralded photon-photon coincidences cannot be distinguished in the heralded intensity correlation if the coherence time is lower than the respective detector jitters~\cite{Bettelli2010, Hockel2011} (see also SI). However, in our case, the electron-heralded photon-photon coincidences remain distinct due to the much better relative photon detection jitters compared to the heralding electron detector (see SI).\\
A discrete version of the intensity correlation function $g^{(2)}_H[q]$ is given by
\begin{align}
    g_H^{(2)}[q,E_\textrm{el}]=\frac{N_\textrm{e,A,B}[q,E_\textrm{el}]N_\textrm{e}(E_\textrm{el})}{N_\textrm{e,A}(E_\textrm{el})N_\textrm{e,B}(E_\textrm{el})}.
\end{align}
The equation substitutes the photon-photon time delay $\tau$ with the number of heralding electrons $q$ by which photon arrivals on detector A are shifted relative to B for cross-correlation. The number of electron-photon coincidences is now counted by only considering a photon detection within a relative time window of $\pm 2.6$ ns (95\% coincidence interval in Fig. \ref{fig:figure2_correlation}c) to a detected electron (for details, see Methods and SI). The case $q=0$ describes coincidences between two photons heralded by the same electron. By definition, $g_\textrm{H}^{(2)}$ approaches unity for large $|q|$.
The resulting intensity correlations are shown in Fig.~\ref{fig:figure4_g2}\textbf{f}. Without energy filtering (top), a small bunching peak is observed with $g^{(2)}_H[0]=1.10$, similar to the unheralded intensity correlation function (cf. Fig.~\ref{fig:figure4_g2}\textbf{d}), but with bin sizes corresponding to the electron detector resolution. Filtering to a single-quantum energy exchange ($m=1$), however, leads to a strong anti-bunching behavior with $g^{(2)}_H(0)$ suppressed to 0.056. The pronounced anti-bunching dip observed for both the time-dependent and discrete intensity correlation functions confirms the high purity of the heralded single-photon state. Two-photon states should, under ideal conditions, also show prominent antibunching. At present, however, technical aspects, including optical losses, electron spectral mixing, and the statistics of the electron source, limit the ultimate purity of the heralded state (see SI). Such technical limitations can be improved upon by time-gated electron sources and enhanced sample designs.\\ 

Quantifying our scheme in terms of electron-heralded photon generation requires a separation between different electron scattering pathways. We observe inelastic electron scattering at the material, as well as the parametric photon generation into broadband spatial optical modes. The unheralded electron spectrum samples a range of coupling constants reaching $g_0^\textrm{EELS}>0.5$, slightly smaller than the values reported for extended planar structures \cite{Adiv2023}. Notably, in the present experiments, a large fraction of generated excitations populate a single broadband optical mode. With a value of $g_0>0.2$, we report the highest coupling of electrons to a detected photonic mode so far.
Reaching the strong-coupling regime by establishing unity coupling constant to a single optical mode~\cite{Kfir2019, Zhao2025, Xie2025, Karnieli2024, Huang2023} will give direct access to a wide class of new phenomena, such as electron-photon entanglement~\cite{Rotunno2023, Rasmussen2024, Henke2025a, Kazakevich2024}, photon-induced electron-electron correlations~\cite{Kfir2021, Kumar2024a} and the generation of more complex non-Gaussian optical states~\cite{Sirotin2024}.\\

In conclusion, we introduce electron-heralded quantum states of light, identifying hallmark features such as antibunching photon statistics and multiphoton-electron correlations. The scheme will facilitate future electron-heralded high-order Fock states, based on further technical optimizations and strong single-mode coupling. Combined with photon or electron state operations, the generation of a wide variety of complex quantum states comes into reach. Broadening the scope of free-electron quantum optics, our results experimentally establish electron beams as a tailored quantum optical resource.

\clearpage

\part*{Methods}
\begin{footnotesize}
\subsection*{Fiber-coupled integrated waveguides}
The photonic chip with a $\text{Si}_3\text{N}_4$ waveguide embedded in $\text{SiO}_2$ 
is fabricated via the Damascene process~\cite{Liu2021} onto a Si substrate. The waveguide (width x height $2.2 \mu\text{m}$ x $780$~nm) has a separation between the bus waveguides, which connect the interaction region with the chip-to-fiber couplers, of around $125\,\mu\text{m}$. The longest path at which electrons can intersect with the waveguide is around $80~\mu\text{m}$ long. However, the curvature of the waveguide, as well as slight angular deviations between the waveguide and the electron beam, reduce the effective interaction length to around $40~\mu\text{m}$. The waveguide has no top cladding, allowing for the optical modes to reach into the vacuum and interact with the electron beam. An ITO (indium tin oxide) layer of $<20$~nm thickness (index of refraction $n=1.99$ at 1550 nm measured at ITO films deposited under similar conditions) on top of the waveguide reduces charging in the presence of an electron beam. The chip is mounted on a custom-made TEM holder~\cite{Henke2021} such that the waveguide sits in parallel to the electron beam. The optical outputs of the symmetrically-built chip are connected to two short pieces of ultra-high numerical aperture (UHNA-7) fiber, which are spliced to single-mode fibers (SMF-28). The single-mode fibers are sent through the hollowed-out TEM holder and a custom vacuum fiber feedthrough. As the photon generation is directional, the optical detection setup is spliced to the SMF-28 fiber, which is connected to the lower output port. 

The main excited optical mode of the waveguide is the quasi-TM\textsubscript{00} mode due to better phase matching. Further modes are the TE\textsubscript{00} and higher-order modes. While the TE\textsubscript{00} mode only shows a small scattering probability in simulations and the phase-matching energy lies in the detection cut-off region, the higher-order waveguide modes are suppressed during their transmission through single-mode optical fibers. Both TM\textsubscript{00} and TE\textsubscript{00} show transmission efficiencies around TM\textsubscript{00} ~20\%, TE\textsubscript{00} ~25\% at the fiber connections. This effectively leads to a spatial single-mode optical detection.

\subsection*{TEM setup}
The experiments are conducted at the Göttingen UTEM (JEOL JEM 2100F) with a Schottky field-emission electron source operated in a continuous (thermal-field) mode. The resulting electron beam shows an energy spread of around 0.6~eV at an electron energy of 100 keV. The current is reduced to around 1.4~pA by adjusting the emitter temperature and by using a condenser aperture of $40~\mu\text{m}$ diameter to not oversaturate the electron detector. The experiments are conducted in low-magnification STEM (scanning transmission electron microscopy) mode, enabling the positioning and scanning of the focused electron beam  (30~nm beam diameter, 1.1~mrad half convergence angle) in the sample plane. The scanning resolution is on the order of 30~nm. All measurements described in the main text are conducted with a stationary electron beam, which is checked and manually readjusted for an optimal combination of photon generation rate and electron transmission onto the detector every 10-30~min. Transmission losses on the electron side arise from charging effects along the waveguide surface, distorting and shifting the electron beam, which reduces incoupling into the 5~mm spectrometer entrance aperture. Post-measurement calculations show an electron transmission efficiency of around 60-70\%. For the scanned data, the electron beam is scanned multiple times across the same region (width x height 200x500nm, 300~ms integration per pixel, pixel size $\sim 40$~nm). The scanned data is used to measure the position-dependent change of the coupling constant. 

\subsection*{Optical setup}
The generated photons are sent through the waveguide and optical fibers to the outside of the TEM holder. The fiber end is spliced to a 50/50 beam splitter (Thorlabs TW1550R5A1), which distributes the photons to two cooled avalanche photodiodes (ID Quantique ID230). As the generated light is measured to be around $1300-1400 \text{nm}$, we assume slightly higher losses and a deviation from the 50/50 coupling ratio at the beam splitter. The count rates indicate a splitting ratio of 52/48. 
The avalanche photodiodes are set to  $>25\%$ detection efficiency and $20\mu\text{s}$ dead time, which leads to 250 and 300 photons/s intrinsic dark counts, respectively, for diode A and B. An evaluation of the transmission and detection efficiency shows an approximate efficiency of detecting a generated photon of 2\% per diode.
The optical spectrum is measured using a 10~nm flat-top bandpass filter (WL Photonics, tunability from 1200 to 1600nm), which is set between the TEM holder and the beam splitter, connected to the optical fibers via FC/APC connectors. The transmission band was moved incrementally between 1270 and 1570nm and the count rate on the detector was measured for every position (see SI). The spectrum shows a central wavelength of 1380 nm and a bandwidth of 100 nm (cf. Fig\ref{fig:figure3_spectrum}\textbf{e}).
The unheralded photon-photon intensity correlation is measured with a combination of superconducting nanowire single-photon detectors (Single Quantum) and a time tagger (Swabian Instruments, Time Tagger Ultra). The nanowire detectors show a detection efficiency $>80\%$ and a time resolution $<20$~ps and, therefore, allow for sharper temporal correlation when operated in combination with a Tagger unit with an RMS jitter of 9~ps. 

\subsection*{Event-based electron detector}
The electrons are separated in energy via a spectrometer (CEOS GmbH, CEFID) and detected on a hybrid pixel detector based on 4 Timepix chips with 256x256 readout pixels each (Amsterdam Scientific Instruments, EM CheeTah T3). The event-based data stream consists of electron hits (0.03-eV energy bin per pixel on the dispersion axis, 1.56-ns timing bin) and synchronized signals from two time-to-digital converters (260-ps timing bin) at which the photon detector signals arrive. Both the electron and photon time measurements are synchronized to the same global readout clock. This allows for the temporal correlation of the electron and photon arrival. Further post-acquisition steps are conducted to improve temporal and energy resolution on the electron side: First, a calibration of the individual pixel response times is done with regard to a mean detector response to enhance time resolution~\cite{Feist2022}. Second, single-particle clustering of the multiple activated pixels/hits (3.4 hits per cluster at 100-keV beam energy, maximum cluster size of 10) gives access to individual electron events. Their localization follows the arithmetic mean position and the earliest arrival time inside a single cluster. A comparison between external current measurements and event counting on the detector shows an average of 3.4 excited pixels per electron. Last, time-dependent jitters of the electron beam position are reduced on a 10-s interval by correcting the position of the zero-loss peak maximum. \\
For the temporal correlation between electrons and photons, every detected electron is tagged with the relative time delay $\tau_\textrm{A/B} = t_\textrm{A/B} - t_\textrm{el}$ to the temporally closest photons on detector A and detector B (absolute propagation timing offsets are subtracted). Hence, the resulting electron-photon-photon correlation data contains the electron kinetic energy $E$ (calibrated by the pixel position) and the (relative) arrival times of all three particles. For scanned data, the electron beam position is also included.\\
A thorough analysis and quantitative fit of event rates for this three-dimensional event histogram ($E, \tau_\textrm{A}, \tau_\textrm{B}$) can be found in the SI. For easier access, individual cuts along or integrals over a specific axis are also analyzed and fitted independently. Generally, this procedure may include heralded multi-indexing of photons within the relevant time interval of 100 ns. For the calculation of the discrete heralded intensity correlation $g_H^{(2)}[q, E_\textrm{el}]$, as well as Fig. \ref{fig:figure2_correlation}d, photons were assigned uniquely, only to the temporarily closest electron.

All data besides the uncorrelated spectrum (Fig.~\ref{fig:figure2_correlation}) is analysed after filtering the electrons to events with a maximum electron-photon time delay of 100~ns, which strongly reduces computation time.

\subsection*{Intensity correlation}
The statistics of photons and the proof for single photon states are often given via the intensity correlation measured in a Hanbury-Brown and Twiss setup in which the signal from two single-photon detectors is cross-correlated and normalized to large time delays~\cite{Signorini2020, Migdall2013}
\begin{align*}
g^{(2)}(\tau)=\frac{P_\textrm{A,B}(\tau)}{P_\textrm{A}P_\textrm{B}}.
\end{align*}
Here, $P_\textrm{A,B}(\tau)$ is the probability of detector A and B being excited at a temporal delay $\tau$ to each other, and $P_\textrm{i}$, i=A,B is the probability of detector excitation A or B.\\
As the electrons do not arrive uniformly, their statistical arrival distribution plays a role in the unheralded photon-photon intensity correlation. To remove the electron influence, an electron-heralded intensity correlation is employed. Heralded intensity correlation follows~\cite{Bocquillon2009}

$$
g^{(2)}(t_\textrm{A}, t_\textrm{B}, t_\textrm{e})=\frac{P_\textrm{A,B,e}(t_\textrm{A}, t_\textrm{B}, t_\textrm{e})P_\textrm{e}}{P_\textrm{A,e}(t_\textrm{A}-t_\textrm{e}) P_\textrm{A,e}(t_\textrm{B}-t_\textrm{e})}.
$$
in which the photon-photon cross-correlation $P_\textrm{A,B,e}(t_\textrm{A}, t_\textrm{B},t_\textrm{e})$ is analysed in combination with the detection of a heralding particle $e$ (see SI for more information). While the numerator $P_\textrm{A,B,e}$ becomes a function of all three particle arrival times, the denominators convert into correlation functions between one photon and the heralding electron $P_\textrm{A/B,e}$ as a function of electron-photon time delays $\tau_\textrm{A/B} = t_\textrm{A/B} - t_\textrm{e}$.  The function is again normalized to large photon-photon time delays, but under the condition that either photon A or photon B is in coincidence with a herald $e$. For electron-heralding intensity correlation, the equation is rewritten in terms of correlation counts $N$ instead of coincidence probabilities, giving rise to Equation \ref{eq:g2-2D}. \\
Due to the addition of a third particle, the heralded intensity correlation is a function of two time delays $\tau_\textrm{A}$ and $\tau_\textrm{B}$. A reduction to one single variable is done in time-averaged intensity correlation~\cite{Bocquillon2009, Bettelli2010}
$$
\overline{g}^{(2)}(\tau_\textrm{A}=0, \tau=\tau_\textrm{B})=\frac{P_\textrm{A,B,e}(\tau)P_\textrm{e}}{P_\textrm{A,e}(0) P_\textrm{A,e}(\tau)}.
$$
in which one time delay $\tau_\textrm{A}$ is set to coincidence. Thus, the second photon B is simultaneously cross-correlated with a photon A and an electron.\\

For a discrete intensity correlation function, two photons are correlated as a function of the number of heralding electrons $q$ by which a photon on detector A is shifted relative to a photon on detector B during cross-correlation. The equation, therefore, compares the number of successful coincidences between two photons correlated to the same electron ($q=0$) to the number of successful coincidences of two photons correlated to two different electrons ($q\neq0$), assuming the electrons came at the same time. A successful coincidence is defined by setting a time interval of 2.9~ns, given by the temporal jitters of the system, in which two particles have to arrive compared to each other. 

The normalization for large $q$ is given by the number of electron-photon coincidences $N_\textrm{e,A/e,B}$ and the number of heralding events $N_\textrm{e}$, giving 
\begin{align}
g^{(2)}[q, E_\textrm{el}]=\frac{N_\textrm{e,A,B}[q, E_\textrm{el}]N_\textrm{e}(E_\textrm{el})}{N_\textrm{e,A}(E_\textrm{el})N_\textrm{e,B}(E_\textrm{el})}.
\end{align}
Due to the random arrival times of electrons in the given setup, $q$ can not be directly converted to a time delay, compared to periodic sources such as pulsed single photon sources using the pulse trigger as a herald.\\

Generally, all (heralded) intensity correlation functions require noise removal in the individual correlation functions to be normalized for large time delays. However, high signal-to-noise ratios in detector counts, as experienced in this experiment, eliminate the necessity for noise removal. All shown intensity correlation measurements were conducted without background removal. Furthermore, the heralding electrons are generally filtered to a defined energy loss to achieve photon number state analysis on the heralded photon side. The requirement of a fixed electron energy loss value is clarified by adding Eel as a further variable in the intensity correlation functions.
\end{footnotesize}

\begin{footnotesize}
\section*{Data availability statement}
The code and data used to produce the figures in this work will be published in the repository \texttt{Zenodo} upon publication of the paper.

\section*{Acknowledgements}
We thank the members of the EBEAM consortium, in particular F. J. García de Abajo, Albert Polman, Mathieu Kociak and Saskia Fiedler, for fruitful discussions.\\
\textbf{Funding Information:} All samples were fabricated in the Center of MicroNanoTechnology (CMi) at EPFL. This material is based upon work supported by the Air Force Office of Scientific Research under award number FA9550-15-1-0250. All experiments were carried out at the Göttingen UTEM Lab, which was funded by the Max Planck Society, the Deutsche Forschungsgemeinschaft (DFG, German Research Foundation) through 432680300/SFB 1456 (project C01) and the Gottfried Wilhelm Leibniz program, and the European Union’s Horizon 2020 research and innovation program under grant agreement No. 101017720 (FET-Proactive EBEAM). Y.Y. acknowledges support from the EU H2020 research and innovation program under the Marie Sklodowska-Curie IF grant agreement No. 101033593 (SEPhIM). \\
\textbf{Author contribution:} A.S.R. and Y.Y. designed the photonic chip device. Z.Q. optimized the coating and fabricated the device together with R.N.W.. Y.Y. optically characterized the chip and A.S.R. packaged it. J.W.H. designed the TEM sample mount. G.A. built the optical setup and performed the TEM experiments. A.F. implemented the clustering and time-tagging of the event-based data, supported by R.H.. The data was analysed by G.A. and A.F. supported by G.H.. G.H., J.W.H., G.A. and H.J. performed the simulations and G.H. devised the theory section. The study was planned and directed by C.R. and T.J.K. The manuscript was written by G.A., A.F., G.H., and C.R., after discussions with and input from all authors.

\noindent\textbf{Additional information:}
Correspondence and requests for materials should be addressed to A.F., Y.Y., T.J.K. and C.R.\\
(\texttt{armin.feist@mpinat.mpg.de, yujia.yang@epfl.ch, tobias.kippenberg@epfl.ch, claus.ropers@mpinat.mpg.de})\\

\noindent\textbf{Competing financial interests:}
The authors declare no competing financial interests.\\

\end{footnotesize}

\bibliographystyle{apsrev4-2}
\bibliography{main_natphys}
\end{document}